\documentclass[12pt]{article}
\usepackage{graphicx}

\begin{document}
\begin{center}
\textbf{SLIGHTLY GENERALIZED MAXWELL CLASSICAL ELECTRODYNAMICS
CAN BE APPLIED TO INNERATOMIC PHENOMENA}

\bigskip\

\textbf{V.M.Simulik and I.Yu.Krivsky}

\smallskip\

\textit{Institute of Electron Physics, Ukrainian National Academy
of Sciences,\\21 Universitetska Str., 88000 Uzhgorod,
Ukraine\\e-mail: sim@iep.uzhgorod.ua}
\end{center}

\begin{abstract}
In order to extend the limits of classical theory application in the
microworld some weak generalization of Maxwell electrodynamics is suggested.
It is shown that weakly generalized classical Maxwell electrodynamics can
describe the intraatomic phenomena with the same success as relativistic
quantum mechanics can do. Group-theoretical grounds for the description of
fermionic states by bosonic system are presented briefly. The advantages of
generalized electrodynamics in intraatomic region in comparison with
standard Maxwell electrodynamics are demonstrated on testing example of
hydrogen atom. We are able to obtain some results which are impossible in
the framework of standard Maxwell electrodynamics. The Sommerfeld - Dirac
formula for the fine structure of the hydrogen atom spectrum is obtained on
the basis of such Maxwell equations without appealing to the Dirac equation.
The Bohr postulates and the Lamb shift are proved to be the consequences of
the equations under consideration. The relationship of the new model with
the Dirac theory is investigated. Possible directions of unification of such
electrodynamics with gravity are mentioned.
\end{abstract}

\section{Introduction}

There is no doubt that the Maxwell classical electrodynamics of macroworld
(without any generalization) is sufficient for the description of
electrodynamical phenomena in macro region. On the other hand it is well
known that for micro phenomena (inneratomic region) the classical Maxwell
electrodynamics (as well as the classical mechanics) cannot work and must be
replaced by quantum theory. Trying to extend the limits of classical
electrodynamics application to the intraatomic region we came to the
conclusion that it is possible by means of generalization of standard
Maxwell classical electrodynamics in the direction of the extencion of its
symmetry. We also use the relationships between the Dirac and Maxwell
equations for these purposes. Furthermore, the relationships between
relativistic quantum mechanics and classical microscopical electrodynamics
of media are investigated. Such relationships are considered here not only
from the mathematical point of view - they are used for construction of
fundamentals of a non-quantum-mechanical model of microworld.

Our non-quantum-mechanical model of microworld is a model of atom on the
basis of slightly generalized Maxwell's equations, i. e. in the framework of
moderately extended classical microscopical electrodynamics of media. This
model is free from probability interpretation and can explain many
intraatomic phenomena by means of classical physics. Despite the fact that
we construct the classical model, for the purposes of such construction we
use essentially the analogy with the Dirac equation and the results which
were achieved on the basis of this equation. Note also that electrodynamics
is considered here in the terms of field strengths without appealing to the
vector potentials as the primary (input) variables of the theory.

The first step in our consideration is the unitary relationship (and wide
range analogy) between the Dirac equation and slightly generalized Maxwell
equations \cite{S1}.

Our second step is the symmetry principle. On the basis of this principle we
introduced in \cite{SK} the most symmetrical form of generalized Maxwell
equations which now can describe both bosons and fermions because they have
(see \cite{SK}) both spin 1 and spin 1/2 symmetries. On the other hand,
namely these equations are unitarily connected with the Dirac equation. So,
we have one more important argument to suggest these equations in order to
describe intraatomic phenomena, i. e. to be the equations of specific
intraatomic classical electrodynamics.

In our third step we refer to Sallhofer, who suggested in \cite{SALL} the
possibility of introduction of interaction with external field as the
interaction with specific media (a new way of introduction of the
interaction into the field equations). Nevertheless, our model of atom (and
of electron) \cite{S1} is essentially different from the Sallhofer's one.

On the basis of these three main ideas we are able to postulate the slightly
generalized Maxwell equations as the equations for intraatomic classical
electrodynamics which may work in atomic, nuclear and particle physics on
the same level of success as the Dirac equation can do. Below we illustrate
it considering hydrogen atom within the classical model.

The interest to the problem of relationship between the Dirac and Maxwell
equations dates back to the time of creation of quantum mechanics \cite{DAR}%
. But the authors of these papers during long time considered only the most
simple example of free and massless Dirac equation. The interest to this
relationship has grown in recent years due to the results \cite{SALL}, where
the investigations of the case $\mathbf{m}_0\neq 0$ and the interaction
potential $\Phi \neq 0$ were started. Another approach was developed in \cite
{DAV}, where the quadratic relations between the fermionic and bosonic
amplitudes were found and used. In our above mentioned papers \cite{S1,SK},
in publications \cite{S2} and herein we consider the linear relations
between the fermionic and bosonic amplitudes. In \cite{S2} we have found the
relationship between the symmetry properties of the Dirac and Maxwell
equations, the complete set of 8 transformations linking these equations,
the relationship between the conservation laws for the electromagnetic and
spinor fields, the relationship between the Lagrangians for these fields.
Here we summarize our previous results and give some new details of the
intraatomic electrodynamics and its application to the hydrogen atom. The
possibilities of unification with gravitation are briefly discussed.

\section{New classical electrodynamical hydrogen atom model}

Consider the slightly generalized Maxwell equations in a medium with
specific form of sources:

\begin{equation}
\begin{array}{c}
\mathrm{curl}\overrightarrow{H}-\partial _0\epsilon \overrightarrow{E}=
\overrightarrow{j}_e, \quad \mathrm{curl}\overrightarrow{E}+\partial _0\mu
\overrightarrow{H}=\overrightarrow{j}_{mag}, \\
\mathrm{div}\epsilon \overrightarrow{E}=\rho _e, \quad \mathrm{div}\mu
\overrightarrow{H}=\rho _{mag,}
\end{array}
\label{f1}
\end{equation}
where $\overrightarrow{E}$ and $\overrightarrow{H}$ are the electromagnetic
field strengths, $\epsilon $ and $\mu $ are the electric and magnetic
permeabilities of the medium being the same as in the electrodynamical
hydrogen atom model of H. Sallhofer \cite{SALL}:

\begin{equation}
\epsilon \left( \overrightarrow{x}\right) =1-\frac{\Phi \left(
\overrightarrow{x}\right) +\mathbf{m}_0}\omega ,\quad \mu \left(
\overrightarrow{x}\right) =1-\frac{\Phi \left( \overrightarrow{x}\right) -%
\mathbf{m}_0}\omega  \label{f2}
\end{equation}
where $\Phi \left( \overrightarrow{x}\right) =-Ze^2/r$ (we use the units: $%
\hbar =c=1$, transition to standard system is fulfilled by the substitution $%
\omega \longrightarrow \hbar \omega ,$ $\mathbf{m}_0\longrightarrow \mathbf{m%
}_0c^2$). The current and charge densities in equations (\ref{f1}) have the
form

\begin{equation}
\begin{array}{c}
\overrightarrow{j_e}=\mathrm{grad}E^0,\quad \overrightarrow{j}_{mag}=-%
\mathrm{grad}H^0, \\
\rho _e=-\epsilon \mu \partial _0E^0+\overrightarrow{E}\mathrm{grad}\epsilon
,\quad \rho _{mag}=-\epsilon \mu \partial _0H^0+\overrightarrow{H}\mathrm{%
grad}\mu ,
\end{array}
\label{f3}
\end{equation}
where $E^0,H^0$ is the pair of functions (two real scalar fields) generating
the densities of gradient-like sources.

One can easily see that equations (\ref{f1}) are not ordinary
electrodynamical equations known from the Maxwell theory. These equations
have the additional terms which can be considered as the magnetic current
and charge densities - in one possible interpretation, or equations (\ref{f1}%
) can be considered as the equations for compound system of electromagnetic $%
\overrightarrow{(E,}$ $\overrightarrow{H)}$ and scalar $E^0,H^0$ fields in
another possible interpretation.

The reasons of our slight generalization of the classical Maxwell
electrodynamics are the following.

1. The standard Maxwell electrodynamics cannot work in intraatomic region
and its equations are not mathematically equivalent to any of quantum
mechanical equations for electron (Schrodinger equation, Dirac equation,
etc...)

2. The existence of direct relationship between the equations (\ref{f1}) and
the Dirac equation for the massive particle in external electromagnetic
field in the stationary case can be applied. Namely these equations were
shown in papers \cite{S1} to be unitary equivalent with such Dirac equation
(see also Sec. 3 below).

3. Equations (\ref{f1}) can be derived from the principle of maximally
possible symmetry - these equations have both spin 1 and spin 1/2 Poincar\'e
symmetries and in the limit of vanishing of the interaction with medium,
where $\epsilon =\mu =1$, they represent \cite{SK} the maximally symmetrical
form of the Maxwell equations. This fact means first of all that from the
group-theoretical point of view of Wigner, Bargmann - Wigner (and of modern
field theory in general) Eqs. (\ref{f1}) can describe both bosons and
fermions (for more details see Sec. 4. below). As a consequence of this fact
one can use these equations particularly for the description of the
electron. On the other hand, this fact means that intraatomic classical
electrodynamics of electron needs further (relatively to that having been
done by Maxwell) symmetrization of Weber - Faraday equations of classical
electromagnetic theory which leads to the maximally symmetrical form (\ref
{f1}). Below we demonstrate the possibilities of the equations (\ref{f1}) in
the description of testing example of hydrogen atom.

Contrary to \cite{S1}, here the equations (\ref{f1}) are solved directly by
means of separation of variables method. It is useful to rewrite these
equations in the mathematically equivalent form where the sources are
maximally simple:

\begin{equation}
\begin{array}{c}
\mathrm{curl}\overrightarrow{H}-\epsilon \partial _0\overrightarrow{E}=%
\overrightarrow{j}_e,\quad \mathrm{curl}\overrightarrow{E}+\mu \partial _0%
\overrightarrow{H}=\overrightarrow{j}_{mag}, \\
\mathrm{div}\overrightarrow{E}=\stackrel{\sim }{\rho _e},\quad \mathrm{div}%
\overrightarrow{H}=\stackrel{\sim }{\rho _{mag}},
\end{array}
\label{f4}
\end{equation}
where

\begin{equation}
\overrightarrow{j_e}=\mathrm{grad}E^0, \quad \overrightarrow{j}_{mag}=-
\mathrm{grad}H^0, \quad \stackrel{\sim }{\rho _e}=-\mu \partial _0E^0, \quad
\stackrel{\sim }{\rho _{mag}}=-\epsilon \partial _0H^0.  \label{f5}
\end{equation}

Consider the stationary solutions of equations (\ref{f4}). Assuming the
harmonic time dependence for the functions $E^0,H^0$

\begin{equation}
\begin{array}{c}
E^0(t,\overrightarrow{x})=E_A^0(\overrightarrow{x})\cos \omega t+E_B^0(%
\overrightarrow{x})\sin \omega t, \\
H^0(t,\overrightarrow{x})=H_A^0(\overrightarrow{x})\cos \omega t+H_B^0(%
\overrightarrow{x})\sin \omega t,
\end{array}
\label{f6}
\end{equation}
we are looking for the solutions of equations (\ref{f4}) in the form

\begin{equation}  \label{f7}
\begin{array}{c}
\overrightarrow{E}(t,\overrightarrow{x})=\overrightarrow{E}_A(
\overrightarrow{x})\cos \omega t+\overrightarrow{E}_B(\overrightarrow{x}
)\sin \omega t, \\
\overrightarrow{H}(t,\overrightarrow{x})=\overrightarrow{H }_A(%
\overrightarrow{x})\cos \omega t+\overrightarrow{H}_B(\overrightarrow{x}
)\sin \omega t.
\end{array}
\end{equation}

For the 16 time-independent amplitudes we obtain the following two nonlinked
subsystems

\begin{equation}
\begin{array}{c}
\mathrm{curl}\overrightarrow{H_A}-\omega \epsilon \overrightarrow{E_B}=%
\mathrm{grad}E_A^0,\quad \mathrm{curl}\overrightarrow{E_B}-\omega \mu
\overrightarrow{H_A}=-\mathrm{grad}H_B^0, \\
\mathrm{div}\overrightarrow{E_B}=\omega \mu E_A^0,\quad \mathrm{\ div}%
\overrightarrow{H}_A=-\omega \epsilon H_B^0,
\end{array}
\label{f8}
\end{equation}
\begin{equation}
\begin{array}{c}
\mathrm{curl}\overrightarrow{H_B}+\omega \epsilon \overrightarrow{E_A}=%
\mathrm{grad}E_B^0,\quad \mathrm{curl}\overrightarrow{E_A}+\omega \mu
\overrightarrow{H_B}=-\mathrm{grad}H_A^0, \\
\mathrm{div}\overrightarrow{E_A}=-\omega \mu E_B^0,\quad \mathrm{div}%
\overrightarrow{H}_B=\omega \epsilon H_A^0.
\end{array}
\label{f9}
\end{equation}

Below we consider only the first subsystem (\ref{f8}). It is quite enough
because the subsystems (\ref{f8}) and (\ref{f9}) are connected with
transformations

\begin{equation}
\begin{array}{c}
E\longrightarrow H,\quad H\longrightarrow -E,\quad \epsilon E\longrightarrow
\mu H,\quad \mu H\longrightarrow -\epsilon E, \\
\epsilon \longrightarrow \mu ,\quad \mu \longrightarrow \epsilon ,
\end{array}
\label{f10}
\end{equation}
which are the generalizations of duality transformation of free
electromagnetic field. Due to this fact the solutions of subsystem (\ref{f9}%
) can be easily obtained from the solutions of subsystem (\ref{f8}).

Furthermore, it is useful to separate equations (\ref{f8}) into the
following subsystems:

\begin{equation}
\begin{array}{c}
\omega \epsilon E_B^3-\partial _1H_A^2+\partial _2H_A^1+\partial _3E_A^0=0,
\\
\omega \epsilon H_B^0+\partial _1H_A^1+\partial _2H_A^2+\partial _3H_A^3=0,
\\
-\omega \mu E_A^0+\partial _1E_B^1+\partial _2E_B^2+\partial _3E_B^3=0, \\
\omega \mu H_A^3-\partial _1E_B^2+\partial _2E_B^1-\partial _3H_B^0=0,
\end{array}
\label{f11}
\end{equation}
and

\begin{equation}
\begin{array}{c}
\omega \epsilon E_B^1-\partial _2H_A^3+\partial _3H_A^2+\partial _1E_A^0=0,
\\
\omega \epsilon E_B^2-\partial _3H_A^1+\partial _1H_A^3+\partial _2E_A^0=0,
\\
\omega \mu H_A^1-\partial _2E_B^3+\partial _3E_B^2-\partial _1H_B^0=0, \\
\omega \mu H_A^2-\partial _3E_B^1+\partial _1E_B^3-\partial _2H_B^0=0.
\end{array}
\label{f12}
\end{equation}

Assuming the spherical symmetry case, when $\Phi (\overrightarrow{x})=\Phi
(r),$ $r\equiv \left| \overrightarrow{x}\right| $, we are making the
transition into the spherical coordinate system and looking for the
solutions in the spherical coordinates in the form

\begin{equation}
\left( E,H\right) \left( \overrightarrow{r}\right) =R_{(E,H)}\left( r\right)
f_{(E,H)}\left( \theta ,\phi \right) ,  \label{f13}
\end{equation}
where $E\equiv \left( E^0,\overrightarrow{E}\right) ,$ $H\equiv \left( H^0,%
\overrightarrow{H}\right) $. We choose for the subsystem (\ref{f11}) the
d'Alembert Ansatz in the form

\begin{equation}  \label{f14}
\begin{array}{c}
\stackrel{-}{E_A^0}=\stackrel{-}{C_{E_4}}R_{H_4}P_{l_{H_4}}^{\stackrel{-}{%
m_4 }}e^{-\stackrel{-}{im_4}\phi }, \\
\stackrel{-}{E_B^k}=\stackrel{-}{C_{E_k}} R_{E_k}P_{l_{E_k}}^{\stackrel{-}{%
m_k}}e^{-\stackrel{-}{im_k}\phi }, \\
\stackrel{-}{H_B^0}=\stackrel{-}{C_{H_4}}R_{E_4}P_{l_{E_4}}^{\stackrel{-}{%
m_4 }}e^{-\stackrel{-}{im_4}\phi }, \quad k=1,2,3. \\
\stackrel{-}{H_A^k}= \stackrel{-}{C_{H_k}}R_{H_k}P_{l_{H_k}}^{\stackrel{-}{%
m_k}}e^{-i\stackrel{-}{ m_k}\phi },
\end{array}
\end{equation}
We use the following representation for $\partial _1,\partial _2,\partial _3$
operators in spherical coordinates
\begin{equation}  \label{f15}
\begin{array}{c}
\partial _1CRP_l^me^{\mp im\phi }= \frac{e^{\mp im\phi }C}{2l+1}\cos \phi
\left( R_{,l+1}P_{l-1}^{m+1}-R_{,-l}P_{l+1}^{m+1}\right) +e^{\mp i(m-1)\phi
}C\frac m{\sin \theta }P_l^m\frac Rr, \\
\partial _2CRP_l^me^{\mp im\phi }= \frac{e^{\mp im\phi }C}{2l+1}\sin \phi
\left( R_{,l+1}P_{l-1}^{m+1}-R_{,-l}P_{l+1}^{m+1}\right) \mp e^{\mp
i(m-1)\phi }C \frac{im}{\sin \theta }P_l^m\frac Rr, \\
\partial _3CRP_l^me^{\mp im\phi }= \frac{e^{\mp im\phi }C}{2l+1}\left(
R_{,l+1}(l+m)P_{l-1}^m+R_{,-l}(l-m+1)P_{l+1}^m\right) .
\end{array}
\end{equation}

Substitutions (\ref{f14}) and (\ref{f15}) together with the assumptions

\begin{equation}
\begin{array}{c}
R_{E_\alpha }=R_E,\quad l_{E_\alpha }=l_E,\quad R_{H_\alpha }=R_H,\quad
l_{H_\alpha }=l_H, \\
\stackrel{-}{m_1}=\stackrel{-}{m_2}=\stackrel{-}{m_3}-1=\stackrel{-}{m_4}%
-1=m, \\
\stackrel{-}{C_{H_1}}=i\stackrel{-}{C_{H_2}},\quad \stackrel{-}{C_{E_2}}=-i%
\stackrel{-}{C_{E_1}},\quad \stackrel{-}{C_{H_4}}=-i\stackrel{-}{C_{E_3}}%
,\quad \stackrel{-}{C_{H_3}}=-i\stackrel{-}{C_{E_4}}, \\
\stackrel{-}{C_{H_2}^I}=\stackrel{-}{C_{E_4}^I}(l_H^I+m+1),\quad \stackrel{-%
}{C_{E_3}^I}=-\stackrel{-}{C_{E_4}^I\equiv \stackrel{-}{C^I}}, \\
\stackrel{-}{C_{E_1}^I}=\stackrel{-}{C_{E_3}^I}(l_E^I-m),\quad
l_H^I=l_E^I-1\equiv l^I, \\
\stackrel{-}{C_{H_2}^{II}}=-\stackrel{-}{C_{E_4}^{II}}(l_H^{II}-m),\quad
\stackrel{-}{C_{E_3}^{II}}=-\stackrel{-}{C_{E_4}^{II}\equiv \stackrel{-}{%
C^{II}}}, \\
\stackrel{-}{C_{E_1}^{II}}=\stackrel{-}{-C_{E_3}^{II}}(l_E^{II}+m+1),\quad
l_H^{II}=l_E^{II}+1\equiv l^{II}
\end{array}
\label{f16}
\end{equation}
into the subsystem (\ref{f11}) guarantee the separation of variables in
these equations and lead to the pair of equations for two radial functions $%
R_E,R_H$ (for the subsystem (\ref{f12}) the procedure is similar):

\begin{equation}
\epsilon \omega R_E^I-R_{H,-l}^I=0,\quad \mu \omega R_H^I+R_{E,l+2}^I=0,
\label{f17}
\end{equation}
\begin{equation}
\epsilon \omega R_E^{II}-R_{H,l+1}^{II}=0,\quad \mu \omega
R_H^{II}+R_{E,-l+1}^{II}=0;\quad \quad R_{,a}\equiv \left( \frac d{dr}+\frac
ar\right) R.  \label{f18}
\end{equation}
For the case $\Phi =-ze^2/r$ the equations (\ref{f17}), (\ref{f18}) coincide
exactly with the radial equations for the hydrogen atom of the Dirac theory
and, therefore, the procedure of their solution is the same as in well-known
monographs on relativistic quantum mechanics. It leads to the well-known
Sommerfeld - Dirac formula for the fine structure of the hydrogen spectrum.
We note only that here the discrete picture of energetic spectrum in the
domain $0<\omega <\mathbf{m}_0c^2$ is guaranteed by the demand for the
solutions of the radial equations (\ref{f17}), (\ref{f18}) to decrease on
infinity (when $r\rightarrow \infty $ ). From the equations (\ref{f17}), (%
\ref{f18}) and this condition the Sommerfeld - Dirac formula

\begin{equation}
\omega =\omega _{nj}^{hyd}=\frac{\mathbf{m}_0c^2}{\hbar \sqrt{1+\frac{\alpha
^2}{\left( n_r+\sqrt{k^2-\alpha ^2}\right) ^2}}}  \label{f19}
\end{equation}
follows, where the notations of the Dirac theory (see, e. g., \cite{BS}) are
used: $n_r=n-k,$ $k=j+1/2,$ $\alpha =e^2/\hbar c$. Let us note once more
that the result (\ref{f19}) is obtained here not from the Dirac equation,
but from the Maxwell equations (\ref{f1}) with sources (\ref{f3}) in the
medium (\ref{f2}).

Substituting (\ref{f16}) into (\ref{f14}) one can easy obtain the angular
part of the hydrogen solutions for the $\overrightarrow{(E},\overrightarrow{%
H,}E^0,H^0)$ field and calculate according to (\ref{f3}) the corresponding
currents and charges. Let us write down the explicit form for the set of
electromagnetic field strengths $\overrightarrow{(E},\overrightarrow{H})$,
which are the hydrogen solutions of equations (\ref{f1}), and also for the
currents and charges generating these field strengths (the complete set of
solutions is represented in \cite{S1}:
\begin{equation}
\begin{array}{c}
\overrightarrow{E^I}=R_E^I\left|
\begin{array}{c}
\begin{array}{c}
\left( -l+m-1\right) P_{l+1}^m\cos m\phi \\
\left( l-m+1\right) P_{l+1}^m\sin m\phi
\end{array}
\\
-P_{l+1}^{m+1}\cos \left( m+1\right) \phi
\end{array}
\right| ,\quad \overrightarrow{H^I}=R_H^I\left|
\begin{array}{c}
\left( l+m+1\right) P_l^m\sin m\phi \\
\left( l+m+1\right) P_l^m\cos m\phi \\
-P_l^{m+1}\sin \left( m+1\right) \phi
\end{array}
\right| , \\
\overrightarrow{j_e^I}=gradR_H^IP_l^{m+1}\cos \left( m+1\right) \phi ,\quad
\overrightarrow{j_{mag}^I}=-gradR_e^IP_{l+1}^{m+1}\sin \left( m+1\right)
\phi , \\
\rho _e^I=-\left( \varepsilon R_E^I\right) _{,l+2}P_l^{m+1}\cos \left(
m+1\right) \phi ,\quad \rho _{mag}^I=-\left( \mu R_H^I\right)
_{,-l}P_{l+1}^{m+1}\sin \left( m+1\right) \phi ,
\end{array}
\label{f20}
\end{equation}
\begin{equation}
\begin{array}{c}
\overrightarrow{E^{II}}=R_E^{II}\left|
\begin{array}{c}
\left( l+m\right) P_{l-1}^m\cos m\phi \\
\left( -l-m\right) P_{l-1}^m\sin m\phi \\
P_{l-1}^{m+1}\cos (m+1)\phi
\end{array}
\right| ,\quad \overrightarrow{H^{II}}=R_H^{II}\left|
\begin{array}{c}
\left( -l+m\right) P_l^m\sin m\phi \\
\left( -l+m\right) P_l^m\cos m\phi \\
-P_l^{m+1}\sin \left( m+1\right) \phi
\end{array}
\right| \\
\overrightarrow{j_e^{II}}=gradR_H^{II}P_l^{m+1}\cos \left( m+1\right) \phi
,\quad \overrightarrow{j_{mag}^{II}}=-gradR_E^{II}P_{l-1}^{m+1}\sin \left(
m+1\right) \phi , \\
\rho _e^{II}=-\left( \varepsilon R_E^{II}\right) _{,-l+1}P_l^{m+1}\cos
\left( m+1\right) \phi ,\quad \rho _{mag}^{II}=-\left( \mu R_H^{II}\right)
_{,l+1}P_{l-1}^{m+1}\sin \left( m+1\right) \phi .
\end{array}
\label{f21}
\end{equation}

In one of the possible interpretations the states of the hydrogen atom are
described by these field strength functions $\overrightarrow{E},%
\overrightarrow{H}$ generated by the corresponding currents and charge
densities.

It is evident from (\ref{f1}) that currents and charges in (\ref{f20}), (\ref
{f21}) are generated by scalar fields $(E^0,H^0)$. Corresponding to (\ref
{f20}), (\ref{f21}) $(E^0,H^0)$ solutions of equations (\ref{f1}) are the
following:

\begin{equation}
\begin{array}{c}
E^{I0}=R_H^IP_l^{m+1}\cos \left( m+1\right) \phi, \quad
H^{I0}=R_E^IP_{l+1}^{m+1}\sin \left( m+1\right) \phi , \\
E^{II0}=R_H^{II}P_l^{m+1}\cos \left( m+1\right) \phi, \quad
H^{II0}=R_E^{II}P_{l-1}^{m+1}\sin \left( m+1\right) \phi .
\end{array}
\label{f22}
\end{equation}

As in quantum theory, the numbers $n=0,1,2,...;$ $j=k-\frac 12=l\mp \frac 12$
$(k=1,2,...,n)$ and $m=-l,-l+1,...,l$ mark both the terms (\ref{f19}) and
the corresponding exponentially decreasing field functions $\overrightarrow{E%
},\overrightarrow{H}$ (and $E^0,H^0$) in (\ref{f20})-(\ref{f22}), i. e. they
mark the different discrete states of the classical electrodynamical field
(and the densities of the currents and charges) which by definitions
describes the corresponding states of hydrogen atom in the model under
consideration.

Note that the radial equations (\ref{f17}), (\ref{f18}) cannot be obtained
if one neglects the sources in equations (\ref{f1}), or one (electric or
magnetic) of these sources. Moreover, in this case there is no solution
effectively concentrated in atomic region.

Now we can show on the basis of this model that the assertions known as
\textit{Bohr's postulates are the consequences of equations (\ref{f1}) and
of their classical interpretation}, i. e. these assertions can be derived
from the model, there is no necessity to postulate them from beyond the
framework of classical physics as it was in Bohr's theory. To derive the
first Bohr's postulate one can calculate the generalized Pointing vector for
the hydrogen solutions (\ref{f20})-(\ref{f22}), i. e. for the compound
system of stationary electromagnetic and scalar fields $\overrightarrow{(E},%
\overrightarrow{H,}E^0,H^0)$

\begin{equation}
\overrightarrow{P}_{gen}=\int d^3x(\overrightarrow{E}\times \overrightarrow{H%
}-\overrightarrow{E}E^0-\overrightarrow{H}H^0).  \label{f23}
\end{equation}
The straightforward calculations show that not only vector (\ref{f23}) is
identically equal to zero but the Pointing vector itself and the term with
scalar fields $(E^0,H^0)$ are also identically equal to zero:

\begin{equation}
\overrightarrow{P}=\int d^3x(\overrightarrow{E}\times \overrightarrow{H}%
)\equiv 0, \quad \int d^3x(\overrightarrow{E}E^0+\overrightarrow{H}%
H^0)\equiv 0.  \label{f24}
\end{equation}
This means that in stationary states hydrogen atom does not emit any
Pointing radiation neither due to the electromagnetic $\overrightarrow{(E},%
\overrightarrow{H})$ field, nor to the scalar $(E^0,H^0)$ field. That is the
mathematical proof of the first Bohr postulate.

The similar calculations of the energy for the same system (in formulae (\ref
{f23})-(\ref{f25}) the functions $\overrightarrow{(E},\overrightarrow{H,}%
E^0,H^0)$ are taken in appropriate physical dimension which is given by the
formula (\ref{f49}) below)

\begin{equation}
P^0=\frac 12\int d^3x\mathcal{E}^{\dagger }\mathcal{E}=\frac 12\int d^3x(%
\overrightarrow{E}^2+\overrightarrow{H}^2+E_0^2+H_0^2)=\omega _{nj}^{hyd}
\label{f25}
\end{equation}
give a constant $\mathbf{W}_{nl}$, depending on $n,l$ (or $n,j$) and
independent of $m$. In our model this constant is to be identified with the
parameter $\omega $ in equations (\ref{f1}) which in the stationary states
of $\overrightarrow{(E},\overrightarrow{H,}E^0,H^0)$ field appears to be
equal to the Sommerfeld - Dirac value $\omega _{nj}^{hyd}$ (\ref{f19}). By
abandoning the $\hbar =c=1$ system and putting arbitrary $"A"$ in equations (%
\ref{f1}) instead of $\hbar $ we obtain final $\omega _{nj}^{hyd}$ with $"A"$
instead of $\hbar $. Then the numerical value of $\hbar $can be obtained by
comparison of $\omega _{nj}^{hyd}$ containing $"A"$ with the experiment.
These facts complete the proof of the second Bohr postulate.

This result means that in this model the Bohr postulates are no longer
postulates, but the direct consequences of the classical electrodynamical
equation (\ref{f1}). Moreover, this means that together with Dirac or
Schrodinger equations we have now the new equation which can be used for
finding the solutions of atomic spectroscopy problems. In contradiction to
the well-known equations of quantum mechanics our equation is the classical
one.

Being aware that few interpretations of quantum mechanics (e.g.: Copenhagen,
statistical, Feynman's, Everett's, transactional, see e. g. \cite{JC})
exist, we are far from thinking that here the interpretation can be the only
one. But the main point is that now the classical interpretation (without
probabilities) is possible.

Today we prefer the following interpretation of hydrogen atom in the
approach, when one considers only the motion of electron in the external
field of the nucleon. In our model the interacting field of the nucleon and
electron is represented by the medium with permeabilities $\epsilon ,\mu $
given by formulae (\ref{f2}). The atomic electron is interpreted as the
stationary electromagnetic-scalar wave $\overrightarrow{(E},\overrightarrow{%
H,}E^0,H^0)$ in medium (\ref{f2}), i.e. as the stationary electromagnetic
wave interacting with massless scalar fields $(E^0,H^0),$ or with complex
massless scalar field $\mathcal{E}^0=E^0-iH^0$ with spin $s=0$. In other
words, the electron can be interpreted as an object having the structure
consisting of a photon and a massless meson with zero spin connected,
probably, with leptonic charge. The role of the massless scalar field is the
following: it generates the densities of electric and magnetic currents and
charges $(\rho ,\overrightarrow{j})$, which are the secondary objects in
such model. The mass is the secondary parameter too. There is no electron as
an input charged massive corpuscle in this model! The mass and the charge of
electron appear only outside such atom according to the law of
electromagnetic induction and its gravitational analogy. That is why no
difficulties of Rutherford - Bohr's model (about different models of atom
see, e. g., \cite{LAK}) of atom are present here! The Bohr postulates are
shown to be the consequences of the model. This interpretation is based on
the hypothesis of bosonic nature of matter (on the speculation of the
bosonic structure of fermions) according to which all the fermions can be
constructed from different bosons (something like new SUSY theory). Of
course, before the experiment intended to observe the structure of electron
and before the registration of massless spinless meson it is only the
hypothesis but based on the mathematics presented here. We note that such
massless spinless boson has many similar features with the Higgs boson and
the transition here from intraatomic (with high symmetry properties) to
macroelectrodynamics (with loss of many symmetries) looks similarly to the
symmetry breakdown mechanism.

The successors of magnetic monopole can try to develop here the monopole
interpretation (see \cite{Lochak} for the review and some new ideas about
monopole) - we note that there are few interesting possibilities of
interpretation but we want to mark first of all the mathematical facts which
are more important than different ways of interpretation.

\section{The unitary relationship between the relativistic quantum mechanics
and classical electrodynamics in medium}

Let us consider the connection between the stationary Maxwell equations

\begin{equation}
\begin{array}{c}
\mathrm{curl}\overrightarrow{H}-\omega \epsilon \overrightarrow{E}=\mathrm{\
grad}E^0,\quad \mathrm{curl}\overrightarrow{E}-\omega \mu \overrightarrow{H}%
=-\mathrm{grad}H^0, \\
\mathrm{div}\overrightarrow{E}=\omega \mu E^0,\quad \mathrm{div}%
\overrightarrow{H}=-\omega \epsilon H^0,
\end{array}
\label{f26}
\end{equation}
which follow from the system (\ref{f8}) after ommitting indices $A,B$, and
the stationary Dirac equation following from the ordinary Dirac equation

\begin{equation}
\left( i\gamma ^\mu \partial _\mu -\mathbf{m}_0+\gamma ^0\Phi \right) \Psi
=0, \quad \Psi \equiv (\Psi ^\alpha ),  \label{f27}
\end{equation}
with $m\neq 0$ and the interaction potential $\Phi \neq 0.$ Assuming the
ordinary time dependence

\begin{equation}
\Psi (x)=\Psi (\overrightarrow{x})e^{-i\omega t}\Longrightarrow \partial
_0\Psi (x)=-i\omega \Psi (x),  \label{f28}
\end{equation}
for the stationary states and using the standard Pauli - Dirac
representation for the $\gamma $ matrices, one obtains the following system
of equations for the components $\Psi ^\alpha $ of the spinor $\Psi $:

\begin{equation}
\begin{array}{c}
-i\omega \epsilon \Psi ^1+(\partial _1-i\partial _2)\Psi ^4+\partial _3\Psi
^3=0, \\
-i\omega \epsilon \Psi ^2+(\partial _1+i\partial _2)\Psi ^3-\partial _3\Psi
^4=0, \\
-i\omega \mu \Psi ^3+(\partial _1-i\partial _2)\Psi ^2+\partial _3\Psi ^1=0,
\\
-i\omega \mu \Psi ^4+(\partial _1+i\partial _2)\Psi ^1-\partial _3\Psi ^2=0,
\end{array}
\label{f29}
\end{equation}
where $\epsilon $ and $\mu $ are the same as in (\ref{f2}). After
substitution in Eqs. (\ref{f29}) instead of $\Psi $ the following column
\begin{equation}
\Psi =\mathrm{column}\left| -H^0+iE^3,-E^2+iE^1,E^0+iH^3,-H^2+iH^1\right| .
\label{f30}
\end{equation}
one obtains Eqs. (\ref{f26}). A complete set of 8 such transformations can
be obtained with the help of the Pauli - Gursey symmetry operators \cite{IBR}
similarly to \cite{S2}.

It is useful to represent the right-hand side of (\ref{f30}) in terms of
components of the following complex function

\begin{equation}  \label{f31}
\mathcal{E}\equiv \left|
\begin{array}{c}
\overrightarrow{\mathcal{E}} \\
\mathcal{E}^0
\end{array}
\right| =\mathrm{column}\left| E^1-iH^1,E^2-iH^2,E^3-iH^3,E^0-iH^0\right| ,
\end{equation}
where $\overrightarrow{\mathcal{E}}=\overrightarrow{E}-i\overrightarrow{H}$
is the well-known form for the electromagnetic field used by Majorana as far
back as near 1930 (see, e.g., \cite{DAR}), and $\mathcal{E}^0=E^0-iH^0$ is a
complex scalar field. In these terms the connection between the spinor and
electromagnetic (together with the scalar) fields has the form

\begin{equation}
\mathcal{E}=W\Psi, \quad \Psi =W^{\dagger }\mathcal{E},  \label{f32}
\end{equation}
where the unitary operator $W$ is the following:

\begin{equation}
W=\left|
\begin{array}{cccc}
0 & iC_{-} & 0 & C_{-} \\
0 & -C_{+} & 0 & iC_{+} \\
iC_{-} & 0 & C_{-} & 0 \\
iC_{+} & 0 & C_{+} & 0
\end{array}
\right|; \quad C_{\mp }\equiv \frac 12(C\mp 1), \quad C\Psi \equiv \Psi
^{*}, \quad C\mathcal{E}\equiv \mathcal{E}^{*}.  \label{f33}
\end{equation}

The unitarity of the operator (\ref{f33}) can be verified easily by taking
into account that the equations

\begin{equation}
\left( AC\right) ^{\dagger }=CA^{\dagger },\quad aC=Ca^{*},\quad \left(
aC\right) ^{*}=Ca  \label{f34}
\end{equation}
hold for an arbitrary matrix $A$ and a complex number $a$. We note that in
the real algebra (i. e. the algebra over the field of real numbers) and in
the Hilbert space of quantum mechanical amplitudes this operator has all
properties of unitarity: $WW^{-1}=W^{-1}W=1,$ $W^{-1}=W^{\dagger }$, plus
linearity.

The operator (\ref{f33}) transforms the stationary Dirac equation

\begin{equation}  \label{f35}
\left[ \left( \omega -\Phi \right) \gamma ^0+i\gamma ^k\partial _k-\mathbf{m}%
_0\right] \Psi \left( \overrightarrow{x}\right) =0
\end{equation}
from the standard representation (the Pauli - Dirac representation) into the
bosonic representation

\begin{equation}  \label{f36}
\left[ \left( \omega -\Phi \right) \widetilde{\gamma }^0+\widetilde{i}%
\widetilde{\gamma }^k\partial _k-\mathbf{m}_0\right] \mathcal{E}\left(
\overrightarrow{x}\right) =0.
\end{equation}

Here the $\widetilde{\gamma }^\mu $ matrices have the following unusual
explicit form

\begin{equation}
\begin{array}{c}
\widetilde{\gamma }^0=\left|
\begin{array}{cccc}
1 & 0 & 0 & 0 \\
0 & 1 & 0 & 0 \\
0 & 0 & 1 & 0 \\
0 & 0 & 0 & -1
\end{array}
\right| C,\quad \widetilde{\gamma }^1=\left|
\begin{array}{cccc}
0 & 0 & i & 0 \\
0 & 0 & 0 & -1 \\
i & 0 & 0 & 0 \\
0 & 1 & 0 & 0
\end{array}
\right| , \\
\widetilde{\gamma }^2=\left|
\begin{array}{cccc}
0 & 0 & 0 & 1 \\
0 & 0 & i & 0 \\
0 & i & 0 & 0 \\
-1 & 0 & 0 & 0
\end{array}
\right| ,\quad \widetilde{\gamma }^3=\left|
\begin{array}{cccc}
-i & 0 & 0 & 0 \\
0 & -i & 0 & 0 \\
0 & 0 & i & 0 \\
0 & 0 & 0 & i
\end{array}
\right|
\end{array}
\label{f37}
\end{equation}
in which $\widetilde{\gamma }^0$ matrix explicitly contains operator $C$ of
complex conjugation. We call the representation (\ref{f37}) the bosonic
representation of the $\gamma $ matrices. In this representation the
imaginary unit $i$ is represented by the $4\times 4$ matrix operator:

\begin{equation}  \label{f38}
\widetilde{i}=\left|
\begin{array}{cccc}
0 & -1 & 0 & 0 \\
1 & 0 & 0 & 0 \\
0 & 0 & 0 & -i \\
0 & 0 & -i & 0
\end{array}
\right| .
\end{equation}

Due to the unitarity of the operator (\ref{f33}) the $\widetilde{\gamma }%
^\mu $ matrices still obey the Clifford-Dirac algebra

\begin{equation}  \label{f39}
\widetilde{\gamma }^\mu \widetilde{\gamma }^\nu +\widetilde{ \gamma }^\nu
\widetilde{\gamma }^\mu =2g^{\mu \nu }
\end{equation}
and have the same Hermitian properties as the Pauli - Dirac $\gamma ^\mu $
matrices:

\begin{equation}
\widetilde{\gamma }^{0\dagger }=\widetilde{\gamma }^0,\quad \widetilde{%
\gamma }^{k\dagger }=-\widetilde{\gamma }^k.  \label{f40}
\end{equation}
Thus, the formulae (\ref{f37}) give indeed an exotic representation of $%
\gamma ^\mu $ matrices.

In the vector-scalar form the equation (\ref{f36}) is as follows

\begin{equation}
-i\mathrm{curl}\overrightarrow{\mathcal{E}}+\left[ \left( \omega -\Phi
\right) C-\mathbf{m}_0\right] \overrightarrow{\mathcal{E}}=-\mathrm{grad}%
\mathcal{E}^0, \quad \mathrm{div}\overrightarrow{\mathcal{E}}=\left[ \left(
\omega -\Phi \right) C+\mathbf{m}_0\right] \mathcal{E}^0.  \label{f41}
\end{equation}

Fulfilling the transition to the common real field strengths according to
the formula $\mathcal{E}=E-iH$ and separating the real and imaginary parts
we obtain equations (\ref{f26}) which are mathematically equivalent to the
equations (\ref{f1}) in stationary case.

We emphasize that the only difference between the equation (\ref{f36}) in
the case of description of fermions and in the case of bosons is the
possibility of choosing $\gamma ^\mu $ matrices: for the case of fermions
these matrices may be chosen in arbitrary form (in each of representations
of Pauli - Dirac, Majorana, Weyl, ...), in the case of the description of
bosons the representation of $\gamma ^\mu $ matrices and their explicit form
\textit{must be fixed} in the form (\ref{f37}). In the case of bosonic
interpretation of Eq. (\ref{f35}) one must fixes the explicit form of $%
\gamma ^\mu $ matrices and of $\Psi $ (\ref{f30}).

The mathematical facts considered here prove the one-to-one correspondence
between the solutions of the stationary Dirac and the stationary Maxwell
equations with 4-currents of gradient-like type. Hence, one can, using (\ref
{f30}), write down the hydrogen solutions of the Maxwell equations (\ref{f1}%
) (or (\ref{f4})) starting from the well-known hydrogen solutions of the
Dirac equation (\ref{f27}), i. e. without special procedure of finding the
solutions of the Maxwell equations, see \cite{S1}.

\section{Some group-theoretical grounds of the model}

Consider briefly the case of absence of interaction of the compound field $%
\overrightarrow{(E},\overrightarrow{H,}E^0,H^0)$ with media, i. e. the case $%
\epsilon =\mu =1$, and the symmetry properties of the corresponding
equations. In this case equations (\ref{f1}) for the system of
electromagnetic and scalar fields $\overrightarrow{(E},\overrightarrow{H,}%
E^0,H^0)$ have the form:
\begin{equation}
\begin{array}{c}
\partial _0\overrightarrow{E}=\mathrm{curl}\overrightarrow{H}-gradE^0,\quad
\partial _0\overrightarrow{H}=-\mathrm{curl}\overrightarrow{E}-gradH^0, \\
\mathrm{div}\overrightarrow{E}=-\partial _0E^0,\quad \mathrm{div}%
\overrightarrow{H}=-\partial _0H^0.
\end{array}
\label{f42}
\end{equation}
The Eqs. (\ref{f42}) are nothing more than the weakly generalized Maxwell
equations ($\epsilon =\mu =1$) with gradient-like electric and magnetic
sources $j_\mu ^e=-\partial _\mu E^0,$ $j_\mu ^{mag}=-\partial _\mu H^0$, i.
e.

\begin{equation}
\overrightarrow{j}_e=-\mathrm{grad}E^0, \quad \overrightarrow{j} _{mag}=-%
\mathrm{grad}H^0, \quad \rho _e=-\partial _0E^0, \quad \rho _{mag}=-\partial
_0H^0.  \label{f43}
\end{equation}

In terms of complex 4-component object $\mathcal{E}=E-iH$ from formula (\ref
{f31}) (and in terms of following complex tensor
\begin{equation}
\mathsf{E}=(\mathsf{E}^{\mu \nu })\equiv \left|
\begin{array}{cccc}
0 & \mathcal{E}^1 & \mathcal{E}^2 & \mathcal{E}^3 \\
-\mathcal{E}^1 & 0 & i\mathcal{E}^3 & -i\mathcal{E}^2 \\
-\mathcal{E}^2 & -i\mathcal{E}^3 & 0 & i\mathcal{E}^1 \\
-\mathcal{E}^3 & i\mathcal{E}^2 & -i\mathcal{E}^1 & 0
\end{array}
\right| )  \label{f44}
\end{equation}
Eqs. (\ref{f42}) can be rewritten in the manifestly covariant forms

\begin{equation}
\partial _\mu \mathcal{E}_\nu -\partial _\nu \mathcal{E}_\mu +i\varepsilon
_{\mu \nu \rho \sigma }\partial ^\rho \mathcal{E}^\sigma =0,\quad \partial
_\mu \mathcal{E}^\mu =0  \label{f45}
\end{equation}
(vector form) and
\begin{equation}
\partial _\nu \mathsf{E}^{\mu \nu }=\partial ^\mu \mathcal{E}^0  \label{f46}
\end{equation}
- tensor-scalar form. It is useful also to consider the following form of
Eqs. (\ref{f42})= (\ref{f45})=(\ref{f46}):

\begin{equation}
(i\partial _0-\overrightarrow{S}\cdot \overrightarrow{p})\overrightarrow{%
\mathcal{E}}-i\mathrm{grad}\mathcal{E}^0=0,\quad \partial _\mu \mathcal{E}%
^\mu =0,  \label{f47}
\end{equation}
where $\overrightarrow{S}\equiv (S^j)$ are the generators of irreducible
representation $D(1)$ of the group $SU(2)$:

\begin{equation}
S^1=\left|
\begin{array}{ccc}
0 & 0 & 0 \\
0 & 0 & -i \\
0 & i & 0
\end{array}
\right| ,\quad S^2=\left|
\begin{array}{ccc}
0 & 0 & i \\
0 & 0 & 0 \\
-i & 0 & 0
\end{array}
\right| ,\quad S^3=\left|
\begin{array}{ccc}
0 & -i & 0 \\
i & 0 & 0 \\
0 & 0 & 0
\end{array}
\right| ,\quad \overrightarrow{S}^2=1(1+1)I.  \label{f48}
\end{equation}

The general solution of Eqs. (\ref{f42})= (\ref{f45})=(\ref{f46})=(\ref{f47}%
) was found in the last references within \cite{S2}, their symmetry
properties were considered in \cite{SK}. This solution was found in the
manifold $(S(R^4)\otimes C^4)^{*}$ of Schwartz's generalized functions
directly by application of Fourier method. In terms of helicity amplitudes $%
c^\mu (\overrightarrow{k})$ this solution has the form

\begin{equation}
\mathcal{E}\left( x\right) =\int \mathrm{d}^3k\sqrt{\frac{2\omega }{\left(
2\pi \right) ^3}}\left\{
\begin{array}{c}
\left[ c^1e_1+c^3\left( e_3+e_4\right) \right] \mathrm{e}^{-ikx}+ \\
\left[ c^{*2}e_1+c^{*4}\left( e_3+e_4\right) \right] \mathrm{e}^{ikx}
\end{array}
\right\} ,\quad \omega \equiv \sqrt{\overrightarrow{k}^2},  \label{f49}
\end{equation}
where $4$-component basis vectors $e_\alpha $ are taken in the form

\begin{equation}
e_1=\left|
\begin{array}{c}
\overrightarrow{e_1} \\
0
\end{array}
\right| ,\quad e_2=\left|
\begin{array}{c}
\overrightarrow{e_2} \\
0
\end{array}
\right| ,\quad e_3=\left|
\begin{array}{c}
\overrightarrow{e_3} \\
0
\end{array}
\right| ,\quad e_4=\left|
\begin{array}{c}
0 \\
1
\end{array}
\right| .  \label{f50}
\end{equation}
Here the $3$-component basis vectors which, without any loss of generality,
can be taken as

\begin{equation}
\overrightarrow{e_1}=\frac 1{\omega \sqrt{2\left( k^1k^1+k^2k^2\right) }%
}\left|
\begin{array}{c}
\omega k^2-ik^1k^3 \\
-\omega k^1-ik^2k^3 \\
i\left( k^1k^1+k^2k^2\right)
\end{array}
\right| ,\quad \overrightarrow{e_2}=\overrightarrow{e_1}^{*},\quad
\overrightarrow{e_3}=\frac{\overrightarrow{k}}\omega ,  \label{f51}
\end{equation}
are the eigenvectors for the quantummechanical helicity operator for the
spin $s=1$.

Note that if the quantities $E^0,H^0$ in Eqs. (\ref{f42}) are some given
functions for which the representation

\begin{equation}
E^0-iH^0=\int \mathrm{d}^3k\sqrt{\frac{2\omega }{\left( 2\pi \right) ^3}}%
\left( c^3\mathrm{e}^{-\mathrm{i}kx}+c^4\mathrm{e}^{\mathrm{i}kx}\right)
\label{f52}
\end{equation}
is valid, then Eqs. (\ref{f42}) are the Maxwell equations with the given
sources, $j_\mu ^e=-\partial _\mu E^0,j_\mu ^{mag}=-\partial _\mu H^0$
(namely these 4 currents we call the gradient-like sources). In this case
the general solution of the Maxwell equations (\ref{f42})= (\ref{f45})=(\ref
{f46})=(\ref{f47}) with the given sources, as follows from (\ref{f49}), has
the form

\begin{equation}
\begin{array}{c}
\overrightarrow{E}\left( x\right) =\int \mathrm{d}^3k\sqrt{\frac \omega
{2\left( 2\pi \right) ^3}}\left( c^1\overrightarrow{e}_1+c^2\overrightarrow{e%
}_2+\alpha \overrightarrow{e}_3\right) \mathrm{e}^{-\mathrm{i}kx}+c.c \\
\overrightarrow{H}\left( x\right) =i\int \mathrm{d}^3k\sqrt{\frac \omega
{2\left( 2\pi \right) ^3}}\left( c^1\overrightarrow{e}_1-c^2\overrightarrow{e%
}_2+\beta \overrightarrow{e}_3\right) \mathrm{e}^{-\mathrm{i}kx}+c.c
\end{array}
\label{f53}
\end{equation}
where the amplitudes of longitudinal waves $\overrightarrow{e}_3\exp \left( -%
\mathrm{i}kx\right) $ are $\alpha =c^3+c^4,$ $\beta =c^3-c^4$ and $c^3,c^4$
are determined by the functions $E^0,H^0$ according to the formula (\ref{f52}%
).

Equations (\ref{f42})= (\ref{f45})=(\ref{f46})=(\ref{f47}) are directly
connected with the free massless Dirac equation

\begin{equation}
i\gamma ^\mu \partial _\mu \Psi (x)=0.  \label{f54}
\end{equation}
There is no reason to appeal here to the stationary case as it was done in
Sec. 3, where the case with nonzero interaction and mass was considered. The
substitution of

\begin{equation}
\psi =\left|
\begin{array}{c}
E^3+iH^0 \\
E^1+iE^2 \\
iH^3+E^0 \\
-H^2+iH^1
\end{array}
\right| =U\mathcal{E},\quad U=\left|
\begin{array}{cccc}
0 & 0 & C_{+} & C_{-} \\
C_{+} & iC_{+} & 0 & 0 \\
0 & 0 & C_{-} & C_{+} \\
C_{-} & iC_{-} & 0 & 0
\end{array}
\right| ,\quad C_{\mp }\equiv \frac 12(C\mp 1),  \label{f55}
\end{equation}
into Dirac equation (\ref{f54}) with $\gamma $ matrices in standard Pauli -
Dirac representation guarantees its transformation into the generalized
Maxwell equations (\ref{f42})= (\ref{f45})=(\ref{f46})=(\ref{f47}). The
complete set of 8 transformations like (\ref{f55}), which relate generalized
Maxwell equations (\ref{f42}) and massless Dirac equation (\ref{f54}), was
found in \cite{S2}. Unitary relationship between the generalized Maxwell
equations (\ref{f45}) and massless Dirac equation (\ref{f54}) was considered
in the way similar to the Sec. 3 and can be found in some our papers from
among the references within \cite{S2}.

Equations (\ref{f45}) (or their another representations (\ref{f42})= (\ref
{f45})=(\ref{f46})=(\ref{f47})) are the maximally symmetrical form of the
generalized Maxwell equations. We consider here representation (\ref{f45})
as an example. The following theorem is valid.

\textbf{Theorem.} \textit{The generalized Maxwell equations (\ref{f45}) are
invariant with respect to the three different transformations, which are
generated by three different representations} $P^V,$ $P^{TS},$ $P^S$ \textit{%
of the Poincar\'e group} $P(1,3)$ \textit{given by the formulae}

\begin{equation}
\begin{array}{c}
\mathcal{E}(x)\rightarrow \mathcal{E}^V(x)=\Lambda \mathcal{E}(\Lambda
^{-1}(x-a)), \\
\mathcal{E}(x)\rightarrow \mathcal{E}^{TS}(x)=F(\Lambda )\mathcal{E}(\Lambda
^{-1}(x-a)), \\
\mathcal{E}(x)\rightarrow \mathcal{E}^S(x)=S(\Lambda )\mathcal{E}(\Lambda
^{-1}(x-a)),
\end{array}
\label{f56}
\end{equation}
\textit{where} $\Lambda $ \textit{is a vector (i. e.} $(\frac 12,\frac 12)$%
\textit{)}, $F(\Lambda )$ \textit{is a tensor-scalar} ($(0,1)\otimes (0,0)$%
\textit{) and} $S(\Lambda )$ \textit{is a spinor representation (}$(0,\frac
12)\otimes (\frac 12,0)$\textit{) of} $SL(2,C)$ \textit{group. This means
that the equations (\ref{f45}) have both spin 1 and spin 1/2 symmetries.}

\textbf{Proof.} Let us write the infinitesimal transformations, following
from (\ref{f56}), in the form
\begin{equation}
\mathcal{E}^{V,TS,S}(x)=(1-a^\rho \partial _\rho -\frac 12\omega ^{\rho
\sigma }j_{\rho \sigma }^{V,TS,S})\mathcal{E}(x).  \label{f57}
\end{equation}
Then the generators of the transformations (\ref{f57}) have the form
\begin{equation}
\partial _\rho =\frac \partial {\partial x^\rho },\quad j_{\rho \sigma
}^{V,TS,S}=x_\rho \partial _\sigma -x_\sigma \partial _\rho +s_{\rho \sigma
}^{V,TS,S},  \label{f58}
\end{equation}
where
\begin{equation}
(s_{\rho \sigma }^V)_\nu ^\mu =\delta _\rho ^\mu g_{\sigma \nu }-\delta
_\sigma ^\mu g_{\rho \nu },\quad s_{\rho \sigma }^V\in \left( \frac 12,\frac
12\right) ,  \label{f59}
\end{equation}
\begin{equation}
s_{\rho \sigma }^{TS}=\left|
\begin{array}{cc}
s_{\rho \sigma }^T & 0 \\
0 & 0
\end{array}
\right| \in (1,0)\oplus (0,0),\quad s_{\rho \sigma }^T=-s_{\sigma \rho
}^T:\quad s_{mn}^T=-i\varepsilon ^{mnj}S^j,\quad s_{0j}^T=S^j,  \label{f60}
\end{equation}
($S^j$ are given by the formula (\ref{f48})) and
\begin{equation}
s_{\rho \sigma }^S=\frac 14[\hat {{\gamma }_\rho },\hat {{\gamma }_\sigma
}],\quad \hat {{\gamma }}=U^{\dagger }\gamma U,  \label{f61}
\end{equation}
(the unitary operator $U$ is given by the formula (\ref{f55}), the explicit
form of $\hat \gamma $ matrices here is essentially different from the
explicit form of the matrices (\ref{f37}) and may be easily found from the
definition in (\ref{f61})). Now the proof of the theorem is reduced to the
verification that all the generators (\ref{f58}) obey the commutation
relations of the $P(1,3)$ group and commute with the operator of the
generalized Maxwell equations (\ref{f42})= (\ref{f45})=(\ref{f46})=(\ref{f47}%
), which can be rewritten in the Dirac form
\begin{equation}
\hat \gamma ^\mu \partial _\mu \mathcal{E}(x)=0  \label{f62}
\end{equation}
(for some details see Ref. \cite{SK}). \textbf{QED.}

This result about the generalized Maxwell equations (\ref{f42})= (\ref{f45}%
)=(\ref{f46})=(\ref{f47}) means the following. From group theoretical point
of view these equations (coinciding with Eqs. (\ref{f1}) in the case $%
\epsilon =\mu =1$) can describe both bosons and fermions. This means that
one has direct group-theoretical grounds to apply these equations for the
description of electron, as it is presented above in Sec. 2.

A distinctive feature of the equation (\ref{f45}) for the system $\mathcal{E}%
=(\overrightarrow{\mathcal{E}},\mathcal{E}^0)$ (i.e. for the system of
interacting irreducible $(0,1)$ and $(0,0)$ fields) is the following. It is
the manifestly covariant equation with minimal number of components, i. e.
the equation without redundant components for this system.

Note that each of the three representations (\ref{f56}) of the $P(1,3)$
group is a local one, because each matrix part of transformations (\ref{f56}%
) (matrices $\Lambda $, $F(\Lambda )$ and $S(\Lambda )$ ) does not depend on
coordinates $x\in R^4$, and, consequently, the generators of (\ref{f56})
belong to the Lie class of operators. Each of the transformations in (\ref
{f56}) may be understood as connected with special relativity
transformations in the space-time $R^4=\{x)$, i. e. with transformations in
the manifold of inertial frame of references.

It follows from the Eqs. (\ref{f45}) that the field $\mathcal{E}=(%
\overrightarrow{\mathcal{E}},\mathcal{E}^0)$ is massless, i. e. $\partial
^\nu \partial _\nu \mathcal{E}^\mu =0$. Therefore it is interesting to note
that neither $P^V$, nor $P^{TS}$ symmetries cannot be extended to the local
conformal $C(1,3)$ symmetry. Only the known spinor $C^S$ representation of $%
C(1,3)$ group obtained from the local $P^S$ representation is the symmetry
group for the generalized Maxwell equations (\ref{f45}). This fact is
understandable: the electromagnetic field $\overrightarrow{\mathcal{E}}=%
\overrightarrow{E}-i\overrightarrow{H}$ obeying Eqs. (\ref{f45}) is not
free, it interacts with the scalar field $\mathcal{E}^0$.

Consider the particular case of standard (non-generalized) Maxwell
equations, namely, the case of equations (\ref{f45}) without magnetic charge
and current densities, i. e. the case when $H^0=0$ but $E^0\neq 0$. The
symmetry properties of such standard equations are strongly restricted in
comparison with the generalized Eqs. (\ref{f45}): they are invariant only
with respect to tensor-scalar (spins 1 and 0) representation of Poincar\'e
group defined by the corresponding representation $(0,1)\otimes (0,0)$ of
proper ortochronous Lorentz group $SL(2,C)$. Another symmetries mentioned in
the theorem are lost for this case. The proof of this assertion follows from
the fact that the vector ($\left( \frac 12,\frac 12\right) $) and the spinor
($\left( 0,\frac 12\right) \oplus \left( \frac 12,0\right) $)
transformations of $\mathcal{E}=(\overrightarrow{\mathcal{E}},\mathcal{E}^0)$
mix the $\mathcal{E}^0$ and $\overrightarrow{\mathcal{E}}$ components of the
field $\mathcal{E}$, and only the tensor-scalar $(0,1)\oplus (0,0)$
transformations do not mix them.

For the free Maxwell equation in vacuum without sources (the case $E^0=H^0=0$%
) the losing of above mentioned symmetries is evident from the same reasons.
Moreover, it is well known that such equations are invariant only with
respect to tensor (spin 1) representations of Poincar\'e and conformal
groups and with respect to dual transformation: $\overrightarrow{E}%
\rightarrow \overrightarrow{H},\overrightarrow{H}\rightarrow -%
\overrightarrow{E}$. We have obtained the extended 32-dimensional Lie
algebra \cite{Ukr} (and the corresponding group) of invariance of free
Maxwell equations, which is isomorphic to $C(1,3)\oplus C(1,3)\oplus dual$
algebra. We were successful to prove it appealing not to Lie class of
symmetry operators but to a more general, namely, to the simplest Lie -
Backlund class of operators. The corresponding generalization of symmetries
of Eqs. (\ref{f45}) presented in the above theorem leads to a wide
246-dimensional Lie algebra in the class of first order Lie - Backlund
operators. Thus, the Maxwell equations (\ref{f45}) with electric and
magnetic gradient-like sources have the maximally possible symmetry
properties among the standard and generalized equations of classical
electrodynamics.

Finally, knowing the operator $U$ (\ref{f55}), it is easy to obtain the
relationship between the amplitudes $a^r(\overrightarrow{k}),$ $b^r(%
\overrightarrow{k})$ determining the well known fermionic solution of the
massless Dirac equation (in Pauli - Dirac representation), and the
amplitudes $c^\alpha (\overrightarrow{k})$, determining the bosonic solution
(\ref{f49}). Corresponding formulae (direct and inverse) related fermionic
and bosonic amplitudes were found in \cite{S2}
\begin{eqnarray}
a^1 &=&\frac 1{2\omega }\left[ i\sqrt{(\omega -k^3)(\omega +k^3)}%
(c^1-c^2)-(\omega -k^3)c^3+(\omega +k^3)c^4\right] ,\quad \omega \equiv
\sqrt{\overrightarrow{k}^2},  \label{f63} \\
a^2 &=&\frac 1{2\omega }\left[ -i(k^1+ik^2)\left( \sqrt{\frac{\omega +k^3}{%
\omega -k^3}}c^1+\sqrt{\frac{\omega -k^3}{\omega +k^3}}c^2\right)
+(k^1+ik^2)(c^3+c^4)\right] ,  \nonumber \\
b^1 &=&\frac 1{2\omega }\left[ i\sqrt{(\omega -k^3)(\omega +k^3)}%
(c^1+c^2)+(\omega +k^3)c^3+(\omega -k^3)c^4\right] ,  \nonumber \\
b^2 &=&\frac 1{2\omega }\left[ i(k^1+ik^2)\left( \sqrt{\frac{\omega -k^3}{%
\omega +k^3}}c^1-\sqrt{\frac{\omega +k^3}{\omega -k^3}}c^2\right)
+(k^1+ik^2)(c^3-c^4)\right] .  \nonumber
\end{eqnarray}
In terms of unitary operator $V$ this formulae have the form:
\begin{equation}
\widehat{a}\equiv \left|
\begin{array}{c}
a^1 \\
a^2 \\
b^1 \\
b^2
\end{array}
\right| =\frac 1{2\omega }\left|
\begin{array}{cccc}
i\sqrt{pq} & -p & -i\sqrt{pq} & q \\
-iz^{*}\sqrt{\frac qp} & z^{*} & -iz^{*}\sqrt{\frac pq} & z^{*} \\
i\sqrt{pq} & q & i\sqrt{pq} & p \\
iz\sqrt{\frac pq} & z & -iz\sqrt{\frac qp} & -z
\end{array}
\right| \cdot \left|
\begin{array}{c}
c^1 \\
c^3 \\
c^2 \\
c^4
\end{array}
\right| =V\cdot \widehat{c},  \label{f64}
\end{equation}
where $p=\omega -k^3,$ $q=\omega +k^3,$ $z=k^1-ik^2,$ $z^{*}=k^1+ik^2,$ $%
\omega \equiv \sqrt{\overrightarrow{k}^2}$. The operator $V$ (the image of
operator $U$ (\ref{f55}) in the space of quantum-mechanical amplitudes $%
\widehat{c}$ and $\widehat{a}$ , i. e. in the rigged Hilbert space $%
S_3^4\subset H\subset S_3^{*4}$, where $S_3^{*4}\equiv (S(R^3)\otimes
C^4)^{*}$ is the space of 4-component generalized Schwartz functions) is
unitary one: $VV^{-1}=V^{-1}V=1,\quad V^{-1}=V^{\dagger },$ plus linearity.

Hence, the fermionic states may be constructed as linear combinations of
bosonic states, namely, of the states of the coupled electromagnetic $%
\overrightarrow{\mathcal{E}}=\overrightarrow{E}-i\overrightarrow{H}$ and
scalar $\mathcal{E}^0=E^0-iH^0$ fields. The inverse relationship between
bosonic and fermionic states is also valid. We prefer the first possibility
which is new (bosonic) realization of the old idea (Thomson, Abraham, etc)
of electromagnetic nature of mass and of material world. Thus, today on the
basis of (\ref{f25}), (\ref{f55}), (\ref{f63}), (\ref{f64}) we may speak
about more general conception of the bosonic field nature of material world.

On the basis of this relationship, in \cite{S2} the relationship between
quantized electromagnetic-scalar and massless spinor field was obtained. The
possibility of both Bose and Fermi quantization types for
electromagnrtic-scalar field (and, inversly, for the Dirac spinor field) was
proved. We will not tauch here the problems of quantization because we are
trying here to demonstrate new possibilities of classical theory.

\section{A brief remark about gravity}

The unified theory of electromagnetic and gravitational phenomena may be
constructed in the approach under consideration in the following way. The
main primary equations again are written as (\ref{f1}) and gravity is
considered as a medium in these equations, i. e. the electric $\epsilon $
and magnetic $\mu $ permeabilities of the medium are some functions of the
gravitational potential $\Phi _{grav}$:

\begin{equation}
\epsilon =\epsilon (\Phi _{grav}),\quad \mu =\mu (\Phi _{grav}).  \label{f65}
\end{equation}

Gravity as a medium may generate all the phenomena which in standard
Einstein's gravity are generated by Riemann geometry. For example, the
refraction of the light beam near a big mass star is a typical medium effect
in such a unified model of electromagnetic and gravitationaal phenomena. The
idea of such consideration consists in the following. The gravitational
interaction between massive objects can be represented as the interaction
with some medium, similarly as here (in Eqs. (\ref{f1})) the electromagnetic
interaction between charged particles is considered.

\section{Brief conclusions}

One of the conclusions of our investigation presented here and in \cite
{SK,S2} is that a field equation itself does not answer the question what
kind of particles (Bose or Fermi) is described by this equation. To answer
this question one needs to find all the representations of the Poincar\'e
group under which the equation is invariant. If more than one such
Poincar\'e representations are found \cite{SK}, including the
representations with integer and half-integer spins, then the given equation
describes both Bose and Fermi particles, and both quantization types (Bose
and Fermi) \cite{S2} of the field function, obeying this equation, satisfy
the microcausality condition. The strict group-theoretical ground of this
assertion is the following \cite{SK}: both slightly generalized Maxwell
equations (\ref{f1}) (with $\epsilon =\mu =1$) and Dirac equation (\ref{f54}%
) (with $\mathbf{m}_0=0,$ $\Phi =0$) are invariant with respect to three
different local representations of Poincar\'e group, namely the standard
spinor, vector and tensor-scalar representations generating by the $(0,\frac
12)\otimes (\frac 12,0),$ $(\frac 12,\frac 12),$ $(0,1)\otimes (0,0)$
representations of the Lorentz SL(2,C) group, respectively.

Now it is clear that only the pair of notions ''equation'' plus ''fixed Bose
or Fermi representation of Poincare group'' answers the question what kind
of particle, boson or fermion, is describing.

So if one fixes the pair ''Dirac equation plus reducible, spins 1 and 0,
representation'' he may describe bosonic system (photon plus boson).

If one fixes another pair ''Dirac equation plus spin 1/2 representation''
one may describe fermions (electron, neutrino, etc.).

If one fixes the pair ''generalized Maxwell equation plus spins 1 and 0
representation'' he may describe bosonic system (photon plus boson).

Finally, if one fixes the pair ''generalized Maxwell equation plus spin 1/2
representation'' one may describe fermions. Namely this last possibility is
under main consideration in this paper.

The simple case $\mathbf{m}_0=0,$ $\Phi =0$ is considered in details in
formulae (\ref{f64}), where it is shown that amplitudes of fermionic states
(or their creation - annihilation operators) are the linear combinations of
amplitudes (or of creation - annihilation operators) of bosonic states. In
this sense our model, where the electron is considered as a compound system
of photon plus massless spinless boson, i. e. the states of electron are the
linear combinations of the states of electromagnetic-scalar field, has the
analogy with modern quark models of hadrons.

In the model of atom under consideration based on the equations of Maxwell's
electrodynamics, not of quantum mechanics, the atomic electron is
interpreted as a classical stationary electromagnetic-scalar wave (the
details of the interpretation see in Sec. 2). That is why this model is
essentially distinguished from the first electrodynamical hydrogen atom
model suggested by Sallhofer, see, e.g., \cite{SALL,LAK}.

A few words can be said about the interpretation of the Dirac $\Psi $
function. As follows from the consideration presented here, e.g., from the
relationship (\ref{f30}), the new interpretation of the Dirac $\Psi $
function can be suggested too: $\Psi $ function is the combination of the
electromagnetic field strengths $\left( \overrightarrow{E},\overrightarrow{H}%
\right) $ and two scalar fields $\left( E^0,H^0\right) $ generating the
electromagnetic sources, i.e. in this case the probability or Copenhagen
interpretation of the function $\Psi $ is not necessary.

In the approach based on the equations (\ref{f1}), it is possible to solve
another stationary problems of atomic physics without any appealing to the
Dirac equation and the probability or Copenhagen interpretation.

Some nonstationary problems, e.g., the problem of transitions between the
stationary states caused by the external perturbation, can be, probably,
solved in terms of the electrodynamical model under consideration similarly
to the solution of this problem on the basis of the stationary Schr\"odinger
equation with the corresponding perturbation.

It is evident from the hydrogen example presented in Sec. 2 that discretness
of the physical system states (and its characteristics such as energy, etc)
may be a consequence not only of quantum systems (Schr\"odinger, Dirac), but
also of the classical (Maxwell) equations for the given system. In the case
under consideration this discretness is caused by the properties of medium,
which are given by the electric and magnetic permeabilities (\ref{f2}).

It is very useful to consider the case of Lamb shift in the approach
presented here. This specific quantum electrodynamical effect (as modern
theory asserts) can be described here in the framework of classical
electrodynamics of media. In order to obtain Lamb shift one must add to $%
\Phi \left( \overrightarrow{x}\right) =-Ze^2/r$ in (\ref{f2}) the
quasipotential (known, e. g., from \cite{HAL}, which follows, of course,
from quantum electrodynamics) and solve the equations (\ref{f1}) for such
medium similarly to the procedure of Sec. 2. Finally one obtains the Lamb
shift correction to the Sommerfeld - Dirac formula (\ref{f19}). Such Lamb
shift can be interpreted as a pure classical electrodynamical effect. It can
be considered here as a consequence of polarization of medium (\ref{f2}) and
not of polarization of such abstract concept as vacuum in quantum
electrodynamics. This brief example demonstrates that our consideration can
essentially extend the limits of classical theory application in microworld,
which was the main purpose of our investigations.

The main conclusion from Sec. 3 is the following. The unitary equivalence
between the stationary Dirac equation and the stationary Maxwell equations
with gradient-like currents and charges in medium (\ref{f2}) gives the
possibility to reformulate all the problems of atomic and nuclear physics
(not only the problem of hydrogen atom description, which here is only an
example of possibilities), which can be solved on the basis of the
stationary Dirac equation, in the language of classical electrodynamical
stationary Maxwell equations. It means that our model in stationary case is
equally successful as the conventional relativistic quantum mechanics.

Thus, the new features which follow from our approach are: (i) the classical
interpretation, (ii) new equation and method in atomic and nuclear physics
based on classical electrodynamics in inneratomic medium like (2), (iii) the
hypothesis of bosonic nature of matter (bosonic structure of fermions), (iv)
extension of the limits of classical theory application in the microworld,
(v) foundations of a unified model of electromagnetic and gravitational
phenomena, in which gravitation is considered as a medium in generalized
equations.

\begin{center}
\textbf{Acknowledgement}
\end{center}

The authors are much grateful to Prof. Hans Sallhofer for multiple
discussions of the details and essential support of the main idea, and to
Prof. Boris Struminsky for useful discussions of alternatives in gravity.

This work is supported by the National Fund of Fundamental Researches of
Ukraine, grant \#F7/458-2001.

\end{document}